\documentclass[aps,reprint,showpacs,floatfix,pre,twocolumn]{revtex4-1}
\usepackage{amsmath}
\usepackage{graphicx}
\usepackage{amssymb}
\usepackage{epsf}
\usepackage{color}
\usepackage{fancyhdr}
\usepackage{rotating}
\usepackage{longtable}
\usepackage{supertabular}
\usepackage{latexsym}
\usepackage{float}
\usepackage{dcolumn}
\usepackage{bm}

\newcommand\be {\begin{equation}}
\newcommand\ee {\end{equation}}
\newcommand\bea {\begin{eqnarray}}
\newcommand\eea {\end{eqnarray}}
\newcommand\n {\nonumber}
\newcommand\bc {\begin{center}}
\newcommand\ec {\end{center}}
\newcommand\bfl{\begin{flushleft}}
\newcommand\efl{\end{flushleft}}
\newcommand\bfr{\begin{flushright}}
\newcommand\efr{\end{flushright}}
\def\bra{\langle}
\def\ket{\rangle}

\begin{document}

\title { Dynamical Properties of Random Field Ising Model}
\author{Suman Sinha}%
\email{suman.sinha.phys@gmail.com}
\affiliation{Department of Physics, University of Calcutta, 
92 Acharya Prafulla Chandra Road, Kolkata 700009, India}
\author{Pradipta Kumar Mandal}
\email{pradipta.mandal@gmail.com}%
\affiliation{Department of Physics, Scottish Church College,
 1 $\&$ 3 Urquhart Square, Kolkata 700006, India}


\begin{abstract}
Extensive Monte Carlo simulations are performed on a two-dimensional random field Ising model.
The purpose of the present work is to study the disorder-induced changes in the
properties of disordered spin systems. The time evolution of the domain growth, the order parameter
and the spin-spin correlation functions are studied in the non equilibrium regime. The dynamical evolution
of the order parameter and the domain growth shows a power law scaling with disorder-dependent exponents.
It is observed that for weak random fields, the two dimensional random field Ising model possesses
long range order.
Except for weak disorder, exchange interaction never wins over pinning 
interaction to establish long range order in the system.
\end{abstract}


\maketitle 

\section{Introduction}
\label{intro}
The random field Ising model (RFIM) belongs to a class of disordered spin models
in which the disorder is coupled to the order parameter of the system. 
A lot has been studied on various aspects of RFIM since Imry and Ma \cite{yiskm}
introduced this model. It has experimental realizations in diluted antiferromagnet
\cite{bkj}.
Its Hamiltonian differs from that of normal Ising model by the addition of a local
random field term which results in a drastic change of its behaviour in both 
equilibrium and non equilibrium situations. In one dimension ($d=1$), the RFIM 
does not order at all \cite{ggdm}. Imry and Ma argued that the random fields
assigned to spins changes the lower critical dimension from $d_l=1$ (pure case)
to $d_l=2$. Later a number of field theoretical calculations \cite{ygpns} suggested
that $d_l=3$. Finally, in 1987, came the exact results by Bricmont and Kupiainen
\cite{jbak} which showed that there is a ferromagnetic phase in three dimension ($3d$).
In 1989, Aizenman and Wehr \cite{majw} provided us with a rigorous proof that there is
no ferromagnetic phase in $2d$ RFIM and thus $d_l=2$. This means that the ground state
is paramagnetic. However, in 1999, Frontera and Vives \cite{cfev} had shown numerical 
signs of a transition in the $2d$ RFIM at $T=0$ below a critical random field strength.
They explained in their paper that the proof by Aizenman and Wehr cannot be misunderstood 
as a proof that ordered phase cannot exist.  We mention in passing that Aizenman, in his recent seminars, claims
that the $2d$ RFIM exhibits a phase transition in the disorder parameter \cite{aizen2}.
Recently, Spasojevic $et. al.$ \cite{sjk} gave numerical evidence that the
$2d$ non equilibrium zero-temperature RFIM exhibits a critical behaviour.
The Hamiltonian for such a system is, in general, given by
\begin{equation}
H=-J\sum_{\langle ij \rangle} s_i s_j + \sum_i \eta_i s_i + H_{\tt ext}\sum_i s_i
\label{genham}
\end{equation}
where $J$ is the coupling constant, conventionally set to unity in the present
work. $\eta_i$ is the quenched random field and $H_{\tt ext}$ is the external magnetic 
field. In the present work the external magnetic field $H_{\tt ext}$ is set to zero.
\begin{figure*}
\begin{center}
\begin{tabular}{cc}
      \resizebox{85mm}{!}{\includegraphics[scale=0.6]{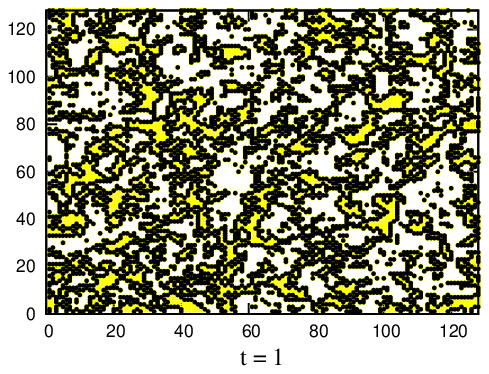}} &
      \resizebox{85mm}{!}{\includegraphics[scale=0.6]{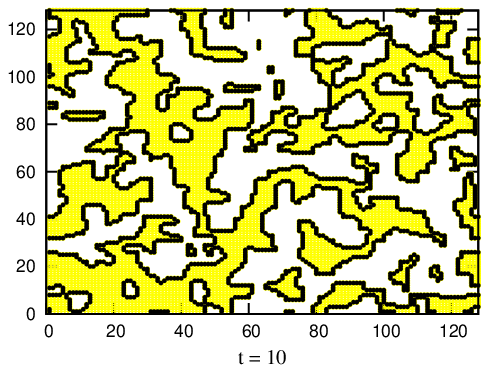}} \\
      \resizebox{85mm}{!}{\includegraphics[scale=0.6]{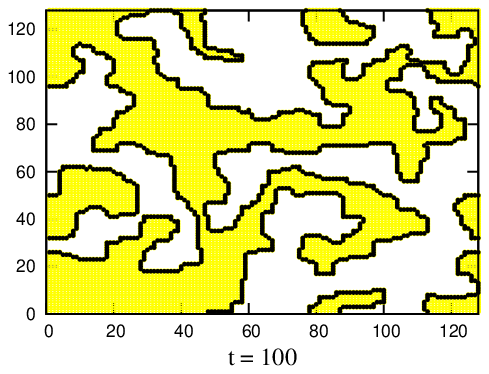}} &
      \resizebox{85mm}{!}{\includegraphics[scale=0.6]{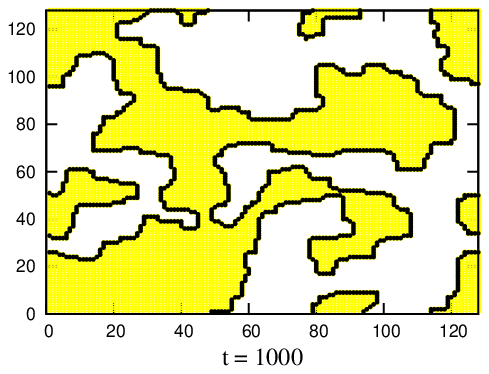}} \\
      \resizebox{85mm}{!}{\includegraphics[scale=0.6]{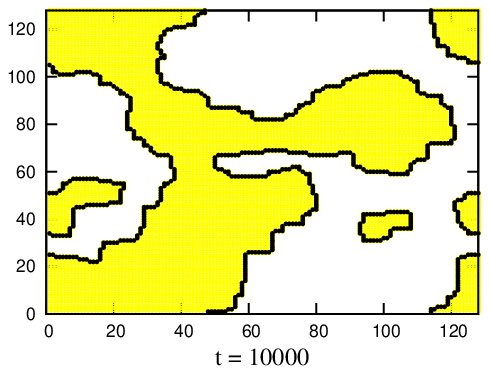}} &
      \resizebox{85mm}{!}{\includegraphics[scale=0.6]{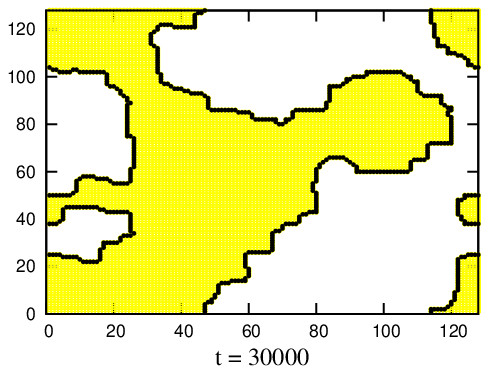}} \\
      \resizebox{85mm}{!}{\includegraphics[scale=0.6]{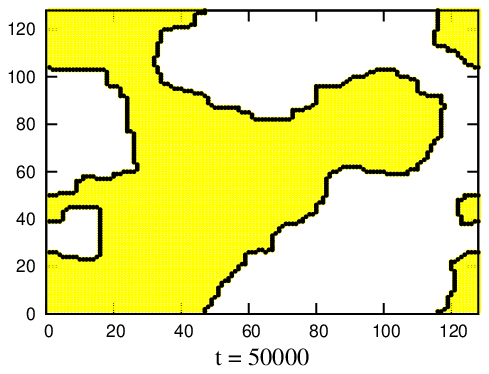}} &
      \resizebox{85mm}{!}{\includegraphics[scale=0.6]{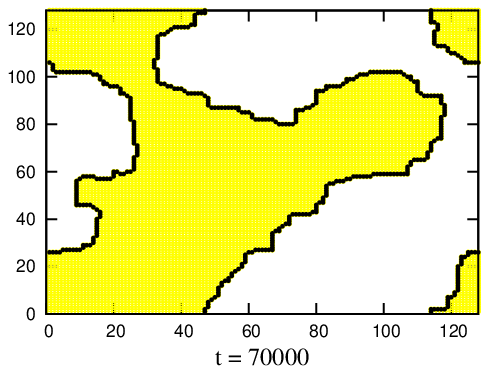}} \\
    \end{tabular}
\end{center}
\caption{(Color online) Time evolution of domains for a $128 \times 128$ system at $T = 0.50$. The 
strength of disorder is $\eta_0 = 1.0$. The yellow (grey) regions indicate domains of up spins.} 
\label{snapshot}
\end{figure*}

The local static random fields give rise to many local minima of the free energy
and the complexity of the random field free energy also gives rise to long 
relaxation times as the system lingers in a succession of local minima on 
its way to the lowest energy state. For the zero field Ising model in two or
more dimensions below the critical temperature $T_c$, there forms domains of 
moderate size in the early time regime, in which all the spins are either up
or down and as time progresses the smallest of these domains shrink and vanish, 
closely followed by the next smallest, until eventually most of the spins on 
the lattice are pointing in the same direction. The reason for this behaviour 
is that the domains of spins possess a surface energy - having a domain wall
cost an energy which increases with the length of the wall, because the spins 
on either side of the wall are pointing in opposite directions. The system can
therefore lower its energy by flipping the spins around the edge of a domain to
make it smaller. Thus the domains ``evaporate'', leaving us most of the spins
either up or down.

However, the story is different for RFIM. In the RFIM, domains still form in 
the ferromagnetic regime, and there is still a surface energy associated with
the domain walls but it is no longer always possible to shrink the domains to
reduce this energy. The random field acting on each spin in the RFIM means that
it has a preferred direction. Furthermore, at some sites there will be a very
large local field $\eta_i$ pointing in one direction, say up direction, which means
that the corresponding spin will really want to point up and it will cost the 
system a great deal of energy if it is pointing down. The acceptance ratio for
flipping this spin contains a factor ${\tt exp} (-2 \beta |\eta_i|)$, which is a 
very small number if $|\eta_i|$ is large. It is said that the domain wall is pinned
by the local field, i.e., it is prevented from moving by an energy barrier 
produced by the large random field. If the domains eventually stop growing, the
system will be in a disordered phase (although the domain size may be very large).
This describes the $d=2$ RFIM.

All these features can be visualized from some snapshots presented in Fig. \ref{snapshot}.
In the early time regime, domains begin to form and grow in size (Fig. \ref{snapshot} (a) - (d)).
This is seen in pure system too.
As time flows, the domains stop growing and the domain wall is pinned (Fig. \ref{snapshot} (e) - (h))
even at a low temperature. The introduction of a local static random field term 
changes the behaviour of the system completely in later time regime and renders the
system in a disordered phase even at a very low temperature.

These disordered spin systems have been an active area of research for quite
some time now \cite{sjk1,tt,vfb,cv,zzh,cadmz,psr}. The purpose of the present work is to study the disorder induced
changes in the properties of these systems, with emphasis on the dynamical 
evolution of the domain growth, the order parameter and the spin-spin correlation 
functions in the non equilibrium region.

In most of the studies on RFIM, the domain size (or the cluster size) was 
determined in terms of the fluctuations of the magnetization \cite{sra,gkgg,gkggs,dcds,askb}. The
fluctuations in magnetization is only a measure of domain size, not the actual
domain size. In this work, we have measured the actual domain size by 
Hoshen-Kopelman algorithm \cite{hk}. The method of determining cluster size
by Hoshen-Kopelman algorithm allows us to make a more accurate study of its
growth with time. Moreover, there are many ways in which the random fields 
could be chosen. In most of the studies, they were chosen to have a 
Gaussian distribution with some finite width $\sigma$, or to have values
randomly $\pm h$ where $h$ is a constant. Although, the interesting 
properties of RFIM are believed to be independent of the exact choice
of the distribution which is a consequence of the phenomenon of universality
\cite{nb}. In the present work, the random fields are chosen from a 
uniform distribution of varying strengths.

The rest of the paper is arranged as follows. Section \ref{simtech} discusses
the computational details and gives the definition of the thermodynamic 
quantities of our interest. Section \ref{rd} presents the results in detail. 
Finally in Section \ref{sc}, we summarize our results.

\section{The model and its simulation}
\label{simtech}
The first term of the Hamiltonian (\ref{genham}) is the usual exchange interaction term while
the second term represents the interaction of the random field with the spin. We call this
term the pinning interaction term since this term is responsible for domain wall pinning.
The simulations are performed on square lattice of sizes ranging from $32 \times 32$ to $512 \times 512$
and at a finite temperature $T=0.50$ which is well below the critical 
temperature of $2d$ zero field Ising model. The temperature is taken sufficiently low to reduce
the thermal fluctuations. Properties of RFIM depend on the competition between the random
fields and the ferromagnetic couplings with the thermal fluctuations serving only to 
renormalize the strengths of these couplings \cite{ywjm}. 
We have chosen Metropolis algorithm \cite{metro} to simulate the system. Metropolis 
algorithm is suitable here because the dynamics is local. Being a single spin flip
dynamics, the Metropolis algorithm is believed to represent the natural way of evolution
of a system, since the acceptance ratio is given by the Boltzmann Probability.
Periodic boundary conditions (PBC) are used in the simulations. 

The sizes of the domains (or clusters)
are determined by Hoshen-Kopelman (HK) algorithm \cite{hk}. The general idea of the HK algorithm
is that we scan through the lattice looking for up spins (or down spins). To each up spin (or down
spin) we assign a label corresponding to the cluster to which the up spin (or the down spin) belongs.
If the up spin (or the down spin) has zero neighbours of same sign, then we assign to it a 
cluster label we have not yet used (it's a new cluster). If the up spin (or the down spin) has 
one neighbour of same sign, then we assign to the current spin the same label as the previous
spin (they are part of the same cluster).  If the up spin (or the down spin) has more than 
one neighbour of same sign, then we choose the lowest-numbered cluster label of the up spins
(or the down spins) to use the label for the current spin. Furthermore, if these neighbouring
spins have different labels, we must make a note that these different labels correspond to the
same cluster. The HK algorithm is a very efficient cluster identification method for two-dimensional
systems. The domain size corresponds to the number of spins enclosed by the boundary of a domain.

The time dependent ensemble average of magnetization, i.e., the order parameter has been defined as
\be
\label{opeqn}
M(\eta_0,t) = \langle \frac{1}{L^2} \mid  \sum_i s_i  \mid \rangle_t
\ee
where $L$ is the linear size of the system and $s_i$ is the spin at the
site $i$. The angular bracket $\langle \cdots \rangle_t$ indicates the ensemble average at 
time $t$.

The time dependent ensemble average of the spin-spin correlation function is defined as
\be
\psi_{ss}(\eta_0, l, t)=\langle \langle s_i s_{i+l}\rangle_l \rangle_t
\label{sseqn}
\ee
where $l$ is the distance of separation between the spins. The angular bracket
$\langle \cdots \rangle_l$ indicates the ensemble average over the distance of
separation between the spins while 
$\langle \cdots \rangle_t$ indicates the same over time $t$.

The thermodynamic quantities of our interest are averaged over $50$ independent
simulations to improve the accuracy and the quality of the results. The simulations
start with a random  spin configurations, characteristic of a high temperature 
phase and then quenched to a low temperature.

\section{Results and Discussions}
\label{rd}
\begin{figure}[!h]
\begin{center}
\resizebox{80mm}{!}{\rotatebox{-90}{\includegraphics[scale=0.6]{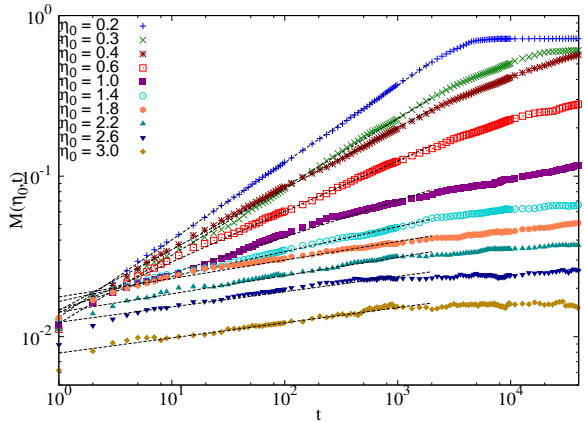}}} 
\end{center}
\caption{(Color online) Plot of the time evolution of order parameter for $L=256$. The dashed lines 
represent the best linear fits according to the scaling law defined 
in Eq. (\ref{opfit})}
\label{opfig}
\end{figure}

\begin{figure}[!h]
\centering
\resizebox{80mm}{!}{\includegraphics[scale=0.6,angle=-90]{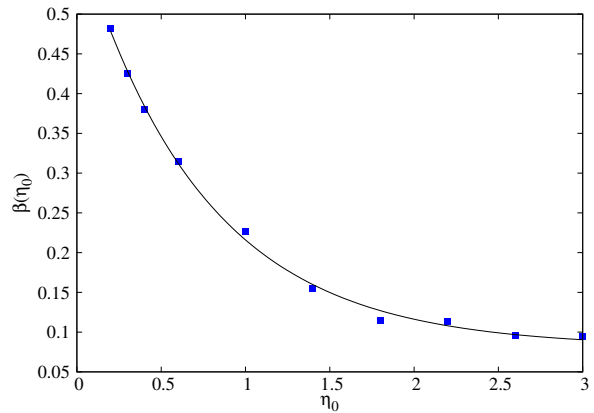}} 
\caption{(Color online) Variation of the exponent $\beta(\eta_0)$ with $\eta_0$. The solid line indicates
the best exponential fit to the data points.}
\label{bvsdel}
\end{figure}

The dynamic evolution of the order parameter defined in Eq. (\ref{opeqn}) 
for different disorder strengths is shown in FIG. \ref{opfig}. 
It is seen from FIG. \ref{opfig} that for weak disorder ($ \eta_0 \le 0.3$)
the system reaches in a steady state after a certain time (here steady state
  means the fluctuations in order parameter is very small with time). For 
weak disorder, the system behaves more or less like a pure system. For larger
disorder, the system takes longer time to reach in a steady state, i.e., the
system relaxes, if at all, very slowly. The nature of the evolution of the
order parameter strongly depends on the strength of the disorder. The early
time behaviour of the dynamic evolution of the order parameter can be
characterized by the following power law behaviour 
  \be
    \label{opfit}
  M(\eta_0,t) \sim t^{\beta(\eta_0)}
  \ee
  where $\beta(\eta_0)$  is a disorder-strength-dependent exponent
  corresponding to the growth of the order parameter. The dependence of the
  power law exponent $\beta(\eta_0)$ on the disorder strength $\eta_0$ is shown
  in FIG. \ref{bvsdel}.
The exponent $\beta(\eta_0)$ falls off exponentially with the strength of the disorder
$\eta_0$ as $\sim {\tt exp}(-\nu \eta_0)$ with $\nu = 1.35 \pm 0.07$. This is quite obvious. 
Two types of interactions, namely, the exchange interaction and the pining interaction are
present in the system. For larger disorder, spin flips are not favoured due to the presence
of the pinning interaction term and consequently the domain walls get pinned, leaving the
system in a disordered phase. As a result, the exponent $\beta(\eta_0)$ falls off sharply with
$\eta_0$.

\begin{figure}[!h]
    \begin{center}
\resizebox{80mm}{!}{\includegraphics[scale=0.6,angle=-90]{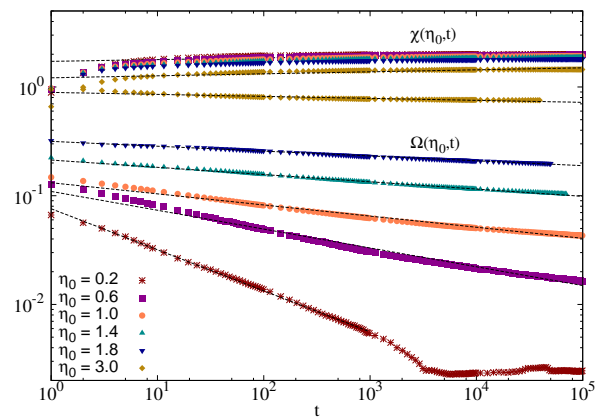}} 
    \end{center}
    \caption{(Color online) The time evolution of the exchange interaction $\chi(\eta_0,t)$
    and the pinning interaction $\Omega(\eta_0,t)$ as defined in
    Eq. (\ref {eieqn}) and Eq. (\ref {pieqn}) respectively for $L=256$.}
    \label{eipifig}
  \end{figure}

What happens is that there is always a competition between the exchange interaction and 
the pining interaction. Therefore it is interesting to observe the time evolution of the
time-dependent ensemble average of the exchange interaction defined as
\be
\chi(\eta_0,t)=\bra \sum_{\bra ij \ket} s_i s_j \ket_t
\label{eieqn}
\ee
as well as the time evolution of the time-dependent ensemble average of the normalized pinning interaction defined as
\be
\Omega(\eta_0,t)=\bra \frac{1}{\eta_0} \bra s_i \eta_i \ket \ket_t
\label{pieqn}
\ee
  where the angular bracket $\langle \cdots \rangle_t$ indicates the ensemble
  average at time $t$. 
  FIG. \ref{eipifig} shows the dynamic evolution of the $\chi(\eta_0,t)$ 
  and the $\Omega(\eta_0,t)$ respectively
and it reveals a striking feature.
  \begin{figure*}
    \begin{center}
\begin{tabular}{cc}
      \resizebox{85mm}{!}{\includegraphics[scale=0.6,angle=-90]{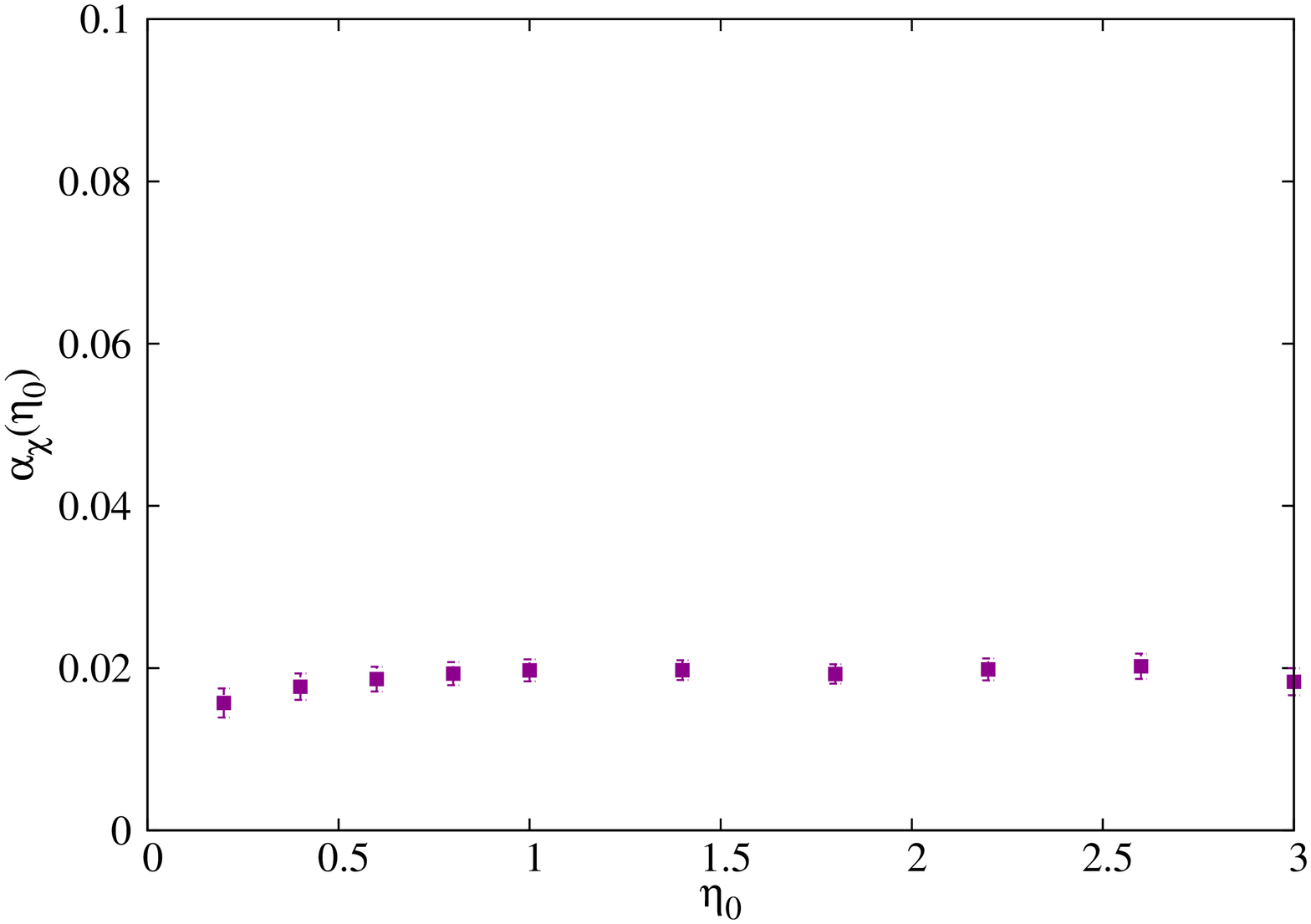}} &
      \resizebox{85mm}{!}{\includegraphics[scale=0.6,angle=-90]{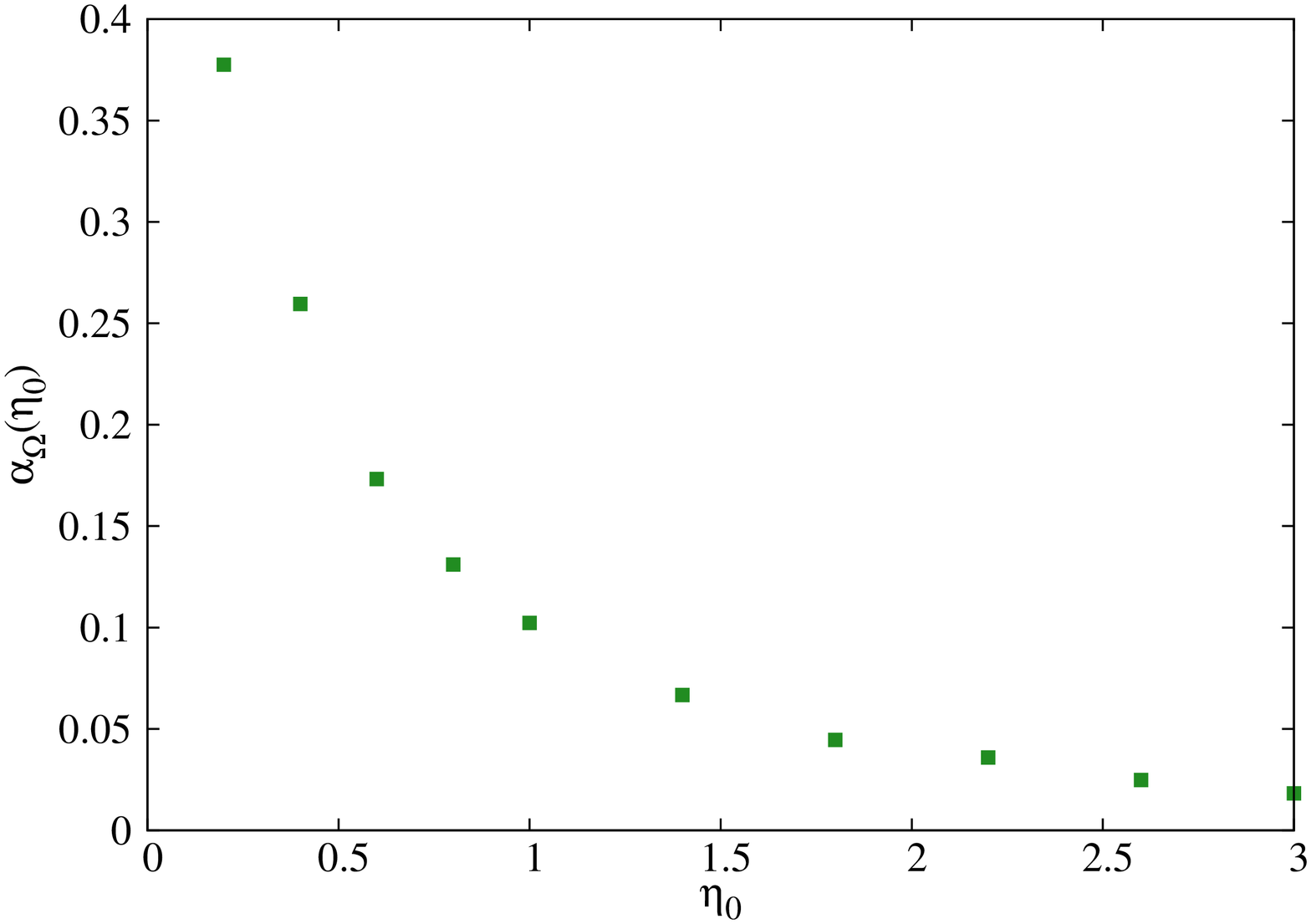}} \\
\end{tabular}
    \end{center}
    \caption{(Color online) Variation of $\alpha_\chi(\eta_0)$ and
      $\alpha_\Omega(\eta_0)$ with $\eta_0$.}
    \label{eipievo}
  \end{figure*}
It is seen that the exchange interaction $\chi(\eta_0,t)$ remains almost same 
with time except at some initial time steps whereas the pinning interaction $\Omega(\eta_0,t)$
  decays more rapidly except for very large disorder strength. The system evolves with time
in such a way that the pining interaction gets minimized. We would like to point out that
this behaviour is characteristic to the RFIM and was not observed earlier to the best 
of our knowledge. 
   For weak disorder, the system achieves the steady state
  with the flow of time resulting in the saturation of the pining interaction and
  the exchange interaction as well. For disorder strengths lying in the intermediate range,
  the decay of pining interaction strongly depends on the strength of the
  disorder. Now a natural question that arises is that how rapidly does the pining
  interaction $\Omega(\eta_0,t)$ decay or how slowly does the exchange interaction
  $\chi(\eta_0,t)$ increase with the strength of
  the random field $\eta_0$. According to the nature of the dynamic
  behaviour seen in FIG. \ref{eipifig}, the power law dependence of the 
 $\chi(\eta_0,t)$ and the $\Omega(\eta_0,t)$ with time has been proposed as
  \bea
  \label{eipieqn}
  \chi(\eta_0,t) \sim t^{\alpha_{\chi}(\eta_0)} \\ \n
  \Omega(\eta_0,t) \sim t^{-\alpha_{\Omega}(\eta_0)}
  \eea
  where $\alpha_{\chi}(\eta_0)$ and $\alpha_{\Omega}(\eta_0)$ are two
  characteristic exponents which determine the nature of growth of the
  exchange interaction and the pining interaction. The variation of these
  exponents 
with $\eta_0$ is shown in FIG. \ref{eipievo}. 
The values of the exponent $\alpha_{\chi}(\eta_0)$ are very small 
and its variation with the strength of the disorder $\eta_0$ is negligible which
indicate that $\alpha_{\chi}(\eta_0)$ is almost independent of $\eta_0$. On the 
other hand, the variation of the exponent $\alpha_{\Omega}(\eta_0)$ with $\eta_0$
is noticeable and $\alpha_{\Omega}(\eta_0)$ decreases rapidly with $\eta_0$
and tends to saturate for very large values of $\eta_0$. These results confirm
our earlier observation on the time evolution of the exchange interaction and
the pinning interaction.

Next we focus our attention on the study of spin-spin correlation functions
defined by Eq. (\ref{sseqn}). The spin-spin correlation functions $\psi_{ss}(\eta_0, l)$
\begin{figure*}
\begin{center}
\begin{tabular}{cc}
      \resizebox{85mm}{!}{\includegraphics[scale=0.6,angle=-90]{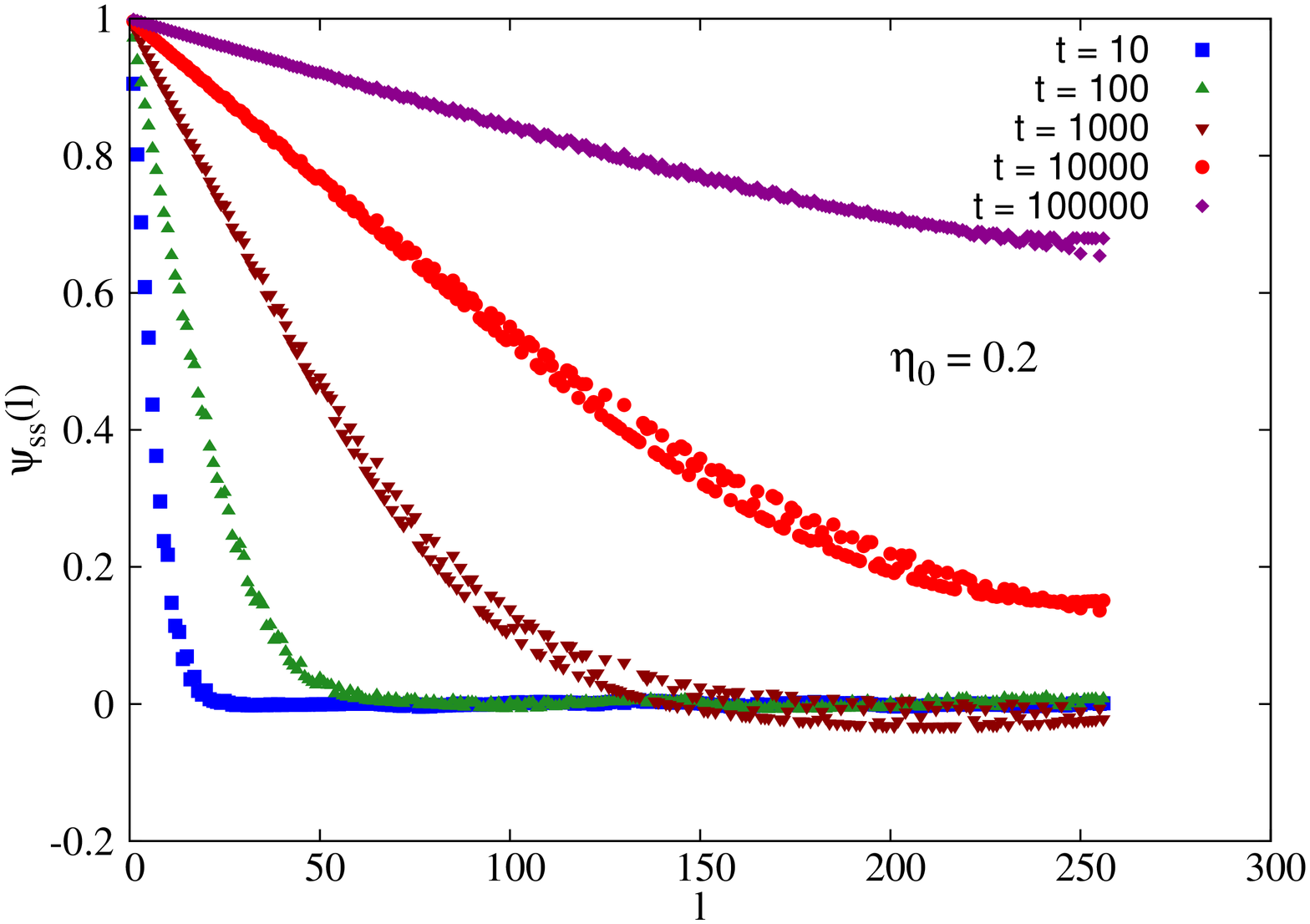}} &
      \resizebox{85mm}{!}{\includegraphics[scale=0.6,angle=-90]{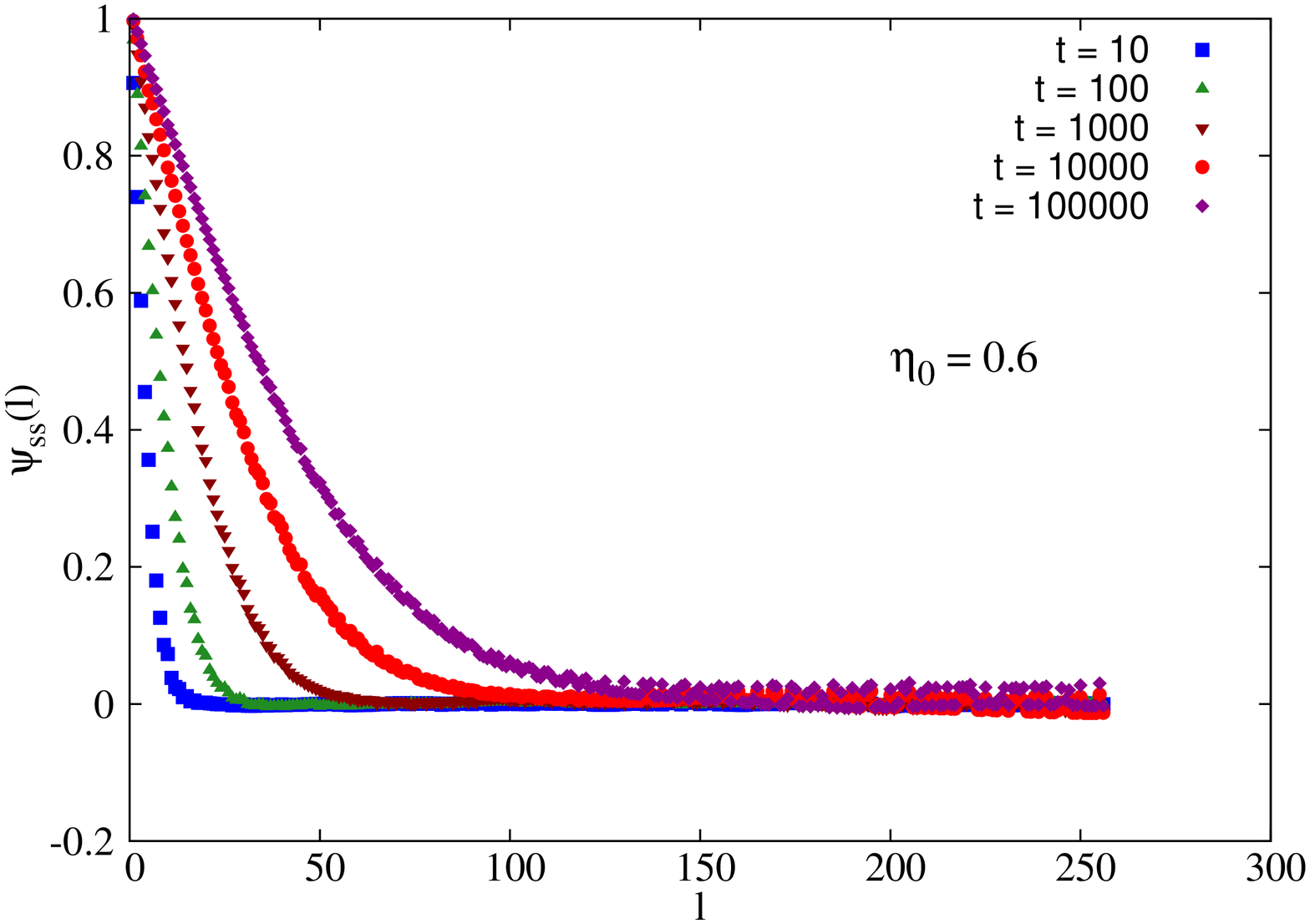}} \\
      \resizebox{85mm}{!}{\includegraphics[scale=0.6,angle=-90]{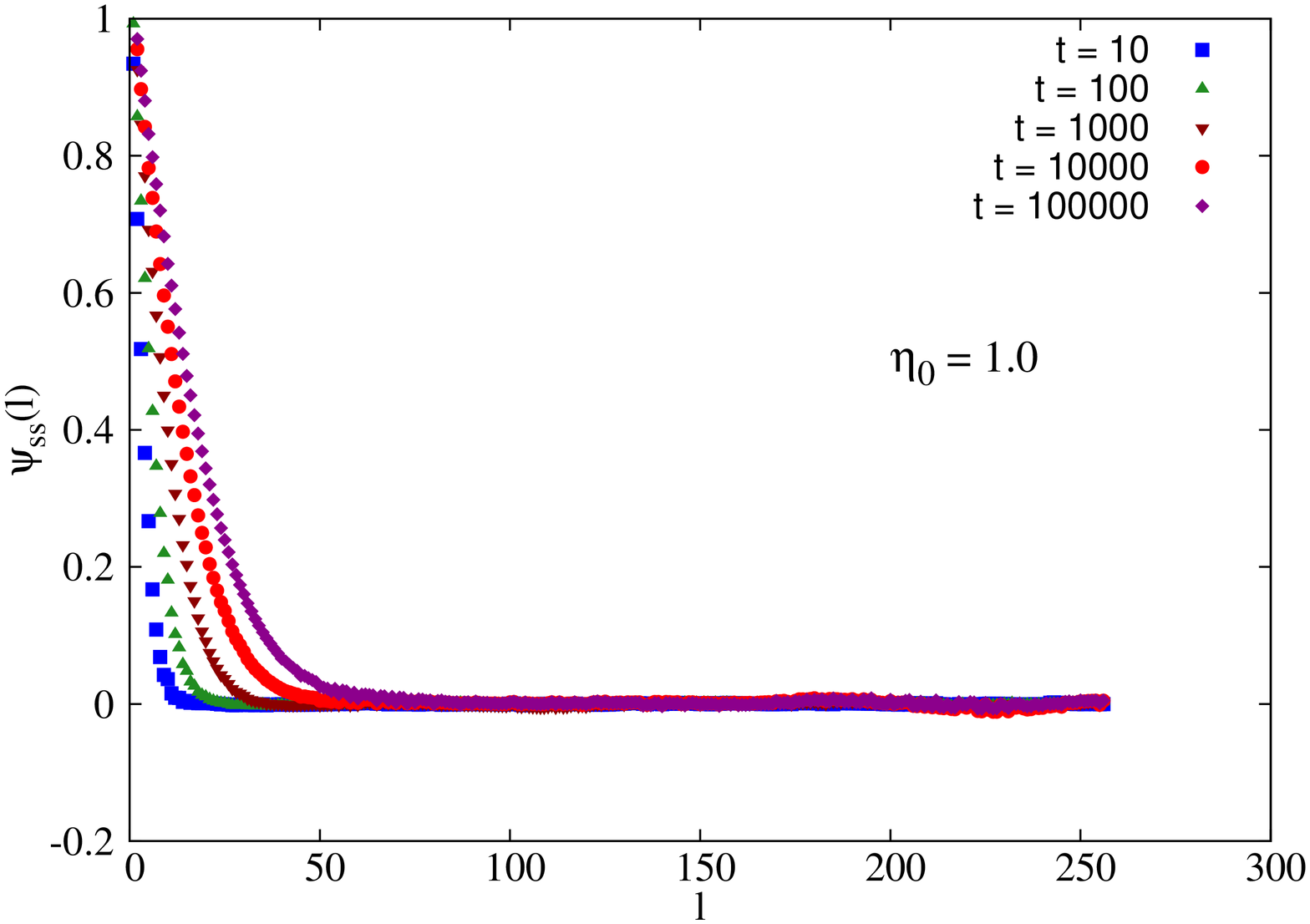}} &
      \resizebox{85mm}{!}{\includegraphics[scale=0.6,angle=-90]{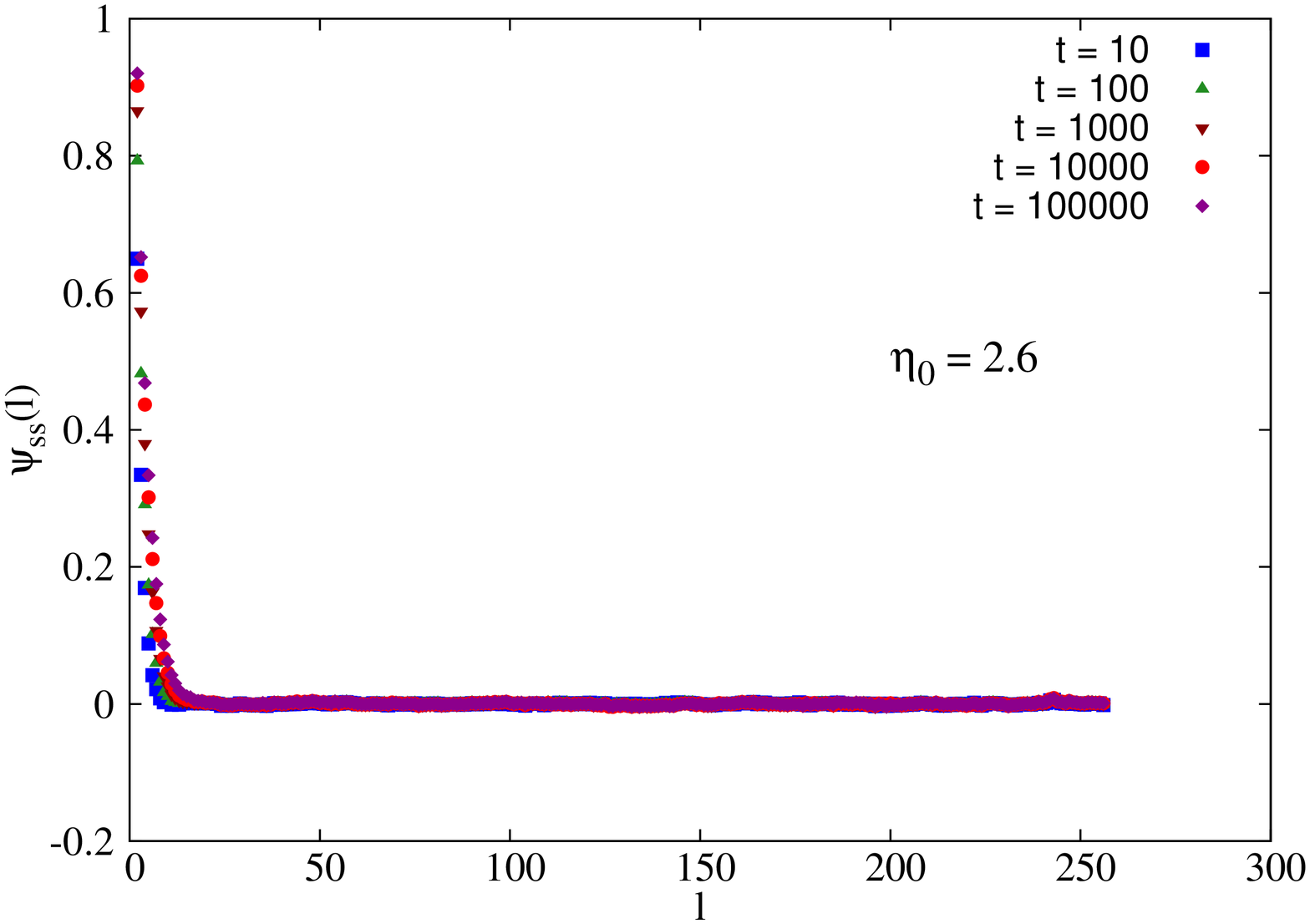}} \\
    \end{tabular}
\end{center}
\caption{(Color online) Plot of $\psi_{ss}(\eta_0, l)$ against $l$ for different disorder strengths
at different times for $L=512$.} 
\label{sscor}
\end{figure*}
for different strengths of disorder at different times $t$ are shown in FIG. \ref{sscor}. 
For weak disorder, when the system is quenched from a high temperature phase to a low
temperature one, long range order is seen to develop at later time regime. With decreasing
strength of the random fields, the ferromagnetic couplings start to dominate over the
random fields, the domains of parallel spins become larger and the system is in a 
ferromagnetic regime. This observation is in agreement with \cite{sa,lkfi} which shows that there
is a critical field strength at which the correlation length becomes divergent. This 
observation also supports the earlier findings \cite{cfev} of a phase transition in the
$2d$ RFIM at $T=0$. Here the thermal fluctuations due to finite temperature serves only
to renormalize the strengths of the random fields and the ferromagnetic couplings. 
For weak disorder, the pinning interaction decays faster with time and then saturates
(see FIG. \ref{eipifig}), which also implies presence of long range order at later time
regime.
This is why, we obtained a large value of the exponents $\beta(\eta_0)$ and $\alpha_\Omega(\eta_0)$
for weak disorder strengths. 
We would also like to point out here that with the increase in system size, it takes longer
time in order for long range order to be established. To check that the presence of long range order for
weak disorder is not a result of finite size effect, we have plotted the order parameter
against time (the number of Monte Carlo steps) for various system sizes, which is shown in
FIG. \ref{lvsmcs}.
  \begin{figure}[!h]
    \begin{center}
\resizebox{80mm}{!}{\includegraphics[scale=0.6]{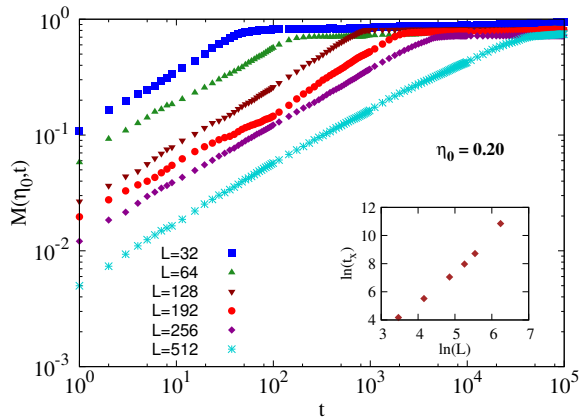}} 
    \end{center}
    \caption{(Color online) Plot of the order parameter against time for various
system sizes at $\eta_0=0.20$.}
    \label{lvsmcs}
  \end{figure}
It is evident from FIG. \ref{lvsmcs} that the number of Monte Carlo steps ($t_x$) required to
achieve the long range order for weak disorder increases with the increase in
system size. The inset of FIG. \ref{lvsmcs} shows the plot of $\ln (t_x)$ against $\ln (L)$.
It may also be noted that for weak disorder, the nature of spin-spin correlation functions
changes from exponential decay to power law decay at late time stage, which suggests the 
possibility of existence of long range order. This observation is in agreement with the 
recent observation of Aizenman \cite{aizen2}.
For strengths of disorder lying in the intermediate range, no
long range order is seen to develop and  short range order is being developed with time in 
the system. 
Except for weak random fields, 
Exchange interaction never wins over pinning interaction to establish long range
order in the system. Due to this short-range order, there forms small domains in the system 
initially and as time elapses, these small domains ``evaporate'' to form a large domain. 
However, the presence of random fields prevents the system from growing into a single domain 
even at a low temperature.
For very large disorder ($\ge 2.6$), neither short range order nor long range order 
prevails in the system and the system remains in a complete disordered phase.
There is always a competition between the ferromagnetic nearest neighbour interaction
$J$ (which favours ordering) and the random field strength $\eta_0$ (which favours
disordering). In the limit of strong random fields, the direction of spins follows 
the direction of random fields.

\begin{figure}[!h]
    \begin{center}
\resizebox{80mm}{!}{\includegraphics[scale=0.6,angle=-90]{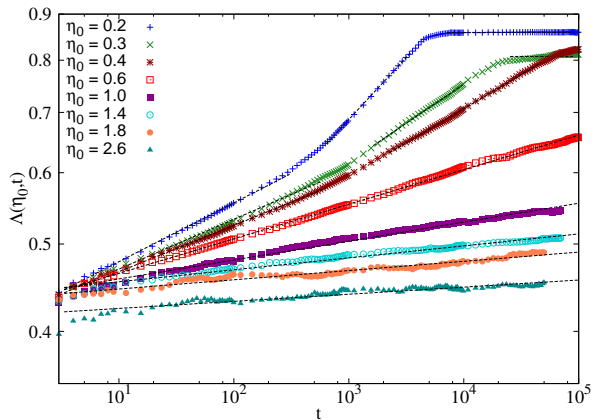}} 
    \end{center}
    \caption{(Color online) Plot of the growth of the largest domain of the system with
      time for $L=256$. The dotted lines represent the best linear fit to the data points.}
    \label{fig6}
  \end{figure}

\begin{figure}[!h]
    \centering
\resizebox{80mm}{!}{\includegraphics[scale=0.6,angle=-90]{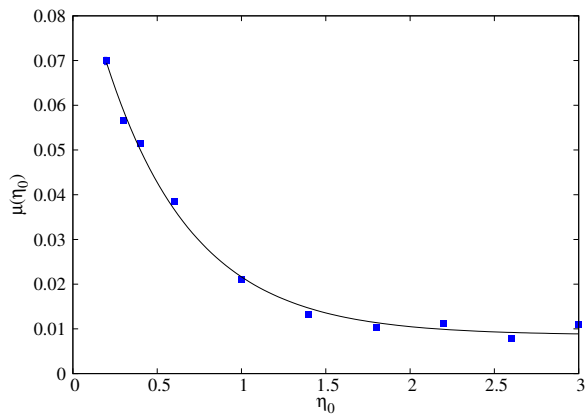}} 
    \caption{(Color online) Plot of the exponents $\mu(\eta_0)$ with disorder strength $\eta_0$.}
    \label{fig6a}
  \end{figure}

Now we present the results of the domain growth with time. Domain growth in quenched
non equilibrium systems is a widely studied topic \cite{bray}. Monte Carlo simulations
have been carried out \cite{shba,gg,pf,cn} to obtain numerically the time evolution 
of these domains, and to compare the results with theory \cite{vil,gf,yb,ba}.
We concentrate on the
growth of the largest domain and it is plotted in FIG. \ref{fig6}.
The growth of the largest domain can be characterized by the following 
power law behaviour :\\
  For $\eta_0 \le 0.3$
  \bea
  \label{dgeqn1}
  \Lambda(\eta_0,t)=t^{\mu(\eta_0)} \,\,\,\,\,\,\,\,\,\,\,\,\,\,\,\,\,\,\,\,\,\, t \ll t_{\times1} \n \\
  \Lambda(\eta_0,t)= t^{\nu(\eta_0)} \,\,\,\,\, t_{\times_1} \ll t \ll t_{\times_2} \n \\
  \Lambda(\eta_0,t) = \eta_0^{\sigma} \,\,\,\,\,\,\,\,\,\,\,\,\,\,\,\,\,\,\,\,\,\,\,\,\,\,\,\, t \gg t_{\times_2}
  \eea
  For  $\eta_0 > 0.3$
  \be
  \label{dgeqn2}
  \Lambda(\eta_0,t)=t^{\mu(\eta_0)}
  \ee
where $\mu(\eta_0)$ is a disorder-strength-dependent exponent corresponding to the 
growth of the largest domain at the early time regime. 
 Recently, Corberi {\it et. al.} \cite{puri} also found that the domain
growth shows a power law scaling with a disorder dependent exponent in preasymptotic regime.
The variation of the exponent 
$\mu(\eta_0)$ with $\eta_0$ is shown in FIG. \ref{fig6a}.
The exponent $\mu(\eta_0)$ falls off exponentially with the strength of disorder $\eta_0$
as $\sim {\tt exp} (-\gamma \eta_0)$ with $\gamma = 1.92 \pm 0.14$. It is to be noted
that although the order parameter exponent and the domain growth exponent fall off 
exponentially with the strength of disorder, the values of the exponents are different and
the fall of $\mu(\eta_0)$ is faster than that of $\beta(\eta_0)$.

\section{Summary and Conclusion}
\label{sc}
We conclude the paper with a summary of our results. This paper has attempted to
consider some aspects of the non equilibrium behaviour of the $2d$ random-field
Ising model numerically at a low temperature. As seen in the preceding Sections, 
the RFIM exhibits a variety of behaviours depending on the strength of the random fields.
The system relaxes with time in presence of two opposite kind of interactions, namely
the exchange interaction and the pinning interaction and we have studied the dynamical
evolution of these two interactions separately. It is seen that the fall of pinning
interaction depends on the strength of the random fields with a power law decay and 
it decays faster for weak random fields. Therefore the dynamical evolution of the order 
parameter should also depend on the strength of the random fields with a power law 
growth, as has been observed. To get an insight of what is happening inside the system,
we have calculated the dynamical spin-spin correlation functions. For weak disorder, the pinning
interaction decays faster and consequently the disordering effect reduces. As a result, the
system is being correlated with time for weak disorder. Our numerical study suggests the
possibility of presence of long range order in the $2d$ RFIM for weak disorder strengths. We are inclined
to comment that the $2d$ RFIM exhibits a phase transition in disorder parameter even at a 
temperature $T > 0$. The transition is manifested by a change of nature of spin-spin correlation
functions from an exponential decay at high disorder strengths to a power law decay at weak disorder
strengths. The thermal fluctuations due to non zero $T$ plays the role only to
renormalize the strengths of both the interactions, although it ceases to be of relevance
at higher temperatures. 
Except for
weak disorder, the exchange interaction never wins over the pinning interaction to 
establish long range order in the system. 
The study of spin-spin correlation functions reveals that 
the $2d$ RFIM shows long-range order,  short-range order and 
no order at all, each of which occurs in a restricted range of random field strength.
We have also measured the largest cluster size by using the Hoshen-Kopelman algorithm.
The behaviours of the dynamical evolution of the largest cluster are consistent with
our previous conclusions.

\section{Acknowledgements}
One of the authors (SS) acknowledges support from the UGC Dr. D. S.
Kothari Post Doctoral Fellowship under grant No. F.4-2/2006(BSR)/13-416/2011(BSR).
SS also thanks Heiko Reiger for many useful discussions. The authors acknowledge 
S. M. Bhattacharjee for a careful reading of the manuscript and a number of comments.
The authors thank the anonymous referee for a number of suggestions in improving the
manuscript.

\end{document}